\documentclass[12pt, a4paper]{article}

\usepackage[utf8]{inputenc}
\usepackage{amsmath, amssymb, amsthm}
\usepackage{geometry}
\usepackage{graphicx}
\usepackage{hyperref}
\usepackage{cite}
\usepackage{xcolor}
\usepackage{bm}
\usepackage{titlesec}
\usepackage{fancyhdr}

\usepackage{authblk}

\usepackage{subfigure}
\usepackage{enumerate}
\usepackage{tikz}
\usetikzlibrary{arrows.meta}
\usepackage{doi}

\usepackage{mdframed}

\newcommand{\N}{\mathbb{N}}

\DeclareMathOperator{\Oh}{\text{Oh}}
\DeclareMathOperator{\Nseg}{N_{\text{seg}}}
\DeclareMathOperator{\Nsamp}{N_{\text{samp}}}

\title{Beyond the Cassie-Baxter Model: New Insights for Predicting Imbibition in Complex Systems}

\author[1]{Mathis Fricke\thanks{Corresponding author: \texttt{fricke@mma.tu-darmstadt.de}}}
\author[2]{Lisanne Gossel}
\author[3]{Joël De Coninck}

\affil[1]{Mathematical Modeling and Analysis, Technical University of Darmstadt, \texttt{fricke@mma.tu-darmstadt.de}}
\affil[2]{Mathematical Modeling and Analysis, Technical University of Darmstadt, \texttt{gossel@mma.tu-darmstadt.de}}
\affil[3]{Transfers, Interfaces and Processes, Université libre de Bruxelles, \texttt{joel.deconinck@umons.ac.be}}

\date{}

\begin{document}

\maketitle

\begin{abstract}
We revisit the classical problem of liquid imbibition in a single pore with spatially varying wettability. Starting from the Lucas-Washburn equation, we derive analytical solutions for the imbibition time (crossing time) in systems where wettability alternates between two materials. For ordered arrangements, we demonstrate that the imbibition speed depends non-trivially on the spatial distribution, with the "more hydrophobic-first" configuration being optimal. For disordered systems, where segment lengths follow a Gaussian distribution, we show that the classical Cassie-Baxter contact angle, originally derived for static wetting, fails to predict the dynamics of capillary-driven flow. To address this, we propose a new weighted harmonic averaging method for the contact angle, which accurately describes the viscous crossing time in such heterogeneous systems. Our findings reveal fundamental insights into the role of wettability heterogeneity in capillary-driven flow, offering a basis for understanding imbibition dynamics in complex heterogeneous systems.\\
\\
The research data and the software supporting this study are openly available at \href{https://doi.org/10.5281/zenodo.14537452}{\textbf{DOI:}10.5281/zenodo.14537452}.
\end{abstract}
\vspace{1em}

\section{Introduction}\label{sec:1-introduction}
Capillary flow through porous media plays a crucial role in applications ranging from material science to biological systems. Examples include the transport of fluids in paper, membranes, and porous filters, as well as fluid uptake in biological systems like trees (see \cite{Shi2020} for an interesting study on synthetic trees). Understanding and controlling capillary-driven flow is also critical in many technologies such as microfluidic devices, additive manufacturing, and the design of advanced materials with tailored wettability. For recent reviews on the wetting of porous systems, see \cite{Cai2021,GambaryanRoisman2014}.\\
\\
In classical models of capillary flow, the Lucas-Washburn equation \cite{Lucas1918,Washburn1921,Bosanquet1923} provides a cornerstone description of spontaneous imbibition, balancing capillary forces with viscous dissipation. While powerful, this equation assumes homogeneous wettability and a static contact angle, which limits its applicability to real-world systems. Many natural and engineered materials exhibit spatial variations in wettability due to heterogeneity in surface properties or chemical composition, significantly influencing the dynamics of imbibition.\\
\\
Previous studies on imbibition in heterogeneous porous media have focused on variations in material properties such as permeability and capillary suction \cite{Suo2018}. These works highlight the critical role of spatial heterogeneity in determining flow dynamics. Similarly, spatial variations in capillary geometry have been studied extensively \cite{Hayek2024}, where heterogeneous radii influence imbibition dynamics through axial nonuniformity. The review \cite{Cai2022} provides a comprehensive discussion of the influence of irregular shapes on the capillary flow. However, the role of spatially varying wettability remains less explored. \\
\\
Recent studies, such as Cai et al. \cite{Cai2021a}, have investigated spatial wettability variations in tree-like networks, revealing their significant impact on spontaneous imbibition dynamics. In contrast, our work focuses on a simpler, single cylindrical capillary with constant geometry but spatially varying wettability. This approach provides a complementary perspective to geometric and material heterogeneity, offering fundamental insights into how wettability distribution affects dynamic capillary-driven flow.\\
\\
A single cylindrical capillary serves as a fundamental \emph{model system} to investigate capillary flow phenomena in more complex porous structures. While real porous materials feature interconnected pores, varying geometries, and surface heterogeneity, the cylindrical capillary provides a tractable system to gain insight into the essential mechanisms governing imbibition. In this work, we extend the classical single-capillary model by allowing for \emph{spatially varying wettability} along the tube, which serves as a simplified representation of material heterogeneity in porous media. Specifically, we explore whether the Cassie-Baxter contact angle \cite{Cassie1944,Cassie1948}, originally derived for static equilibrium, remains a useful concept for predicting the dynamics of capillary-driven flow. By analyzing both ordered and disordered distributions of wettability, we aim to provide new insights into optimizing capillary flow in complex materials.\\
\\
The remainder of this paper is organized as follows: Section \ref{sec:1-1-lucas-washburn} and \ref{sec:1-2-cassie-baxter} briefly revisit the basics of the Lucas-Washburn and Cassie-Baxter equations. In Section~\ref{sec:2-analytical-solution}, we extend the Lucas-Washburn equation to heterogeneous systems and analyze the effect of ordered material distributions on imbibition dynamics. Section~\ref{sec:3-binary-materials} focuses on disordered systems with Gaussian segment distributions, introducing a weighted harmonic averaging method to predict the effective contact angle. Finally, Section~\ref{sec:4-conclusion} concludes the work and outlines future research directions.

\subsection{The classical Lucas-Washburn equation}\label{sec:1-1-lucas-washburn}
We briefly recap the classical Lucas-Washburn equation \cite{Lucas1918,Washburn1921,Bosanquet1923} describing the spontaneous imbibition of liquid into a cylindrical tube of radius $R$. It was originally derived from a simple force balance argument between the capillary driving force and a viscous friction (or dissipation) force. Notably, the forces acting on the liquid column inside the tube are modeled based on simplifying assumptions, which renders the model as an \emph{approximation} to more detailed descriptions based on the two-phase Navier Stokes equations (see, e.g., \cite{Gruending2020b}) or even Molecular Dynamics (see, e.g., \cite{Martic2004}). Nevertheless, the model is well-established in the field and has proven to be compatible with experiments and more detailed numerical simulations if it is applied appropriately (keeping in mind the limitations).
\definecolor{MyBlue}{rgb}{0.9,1.5,1.5}
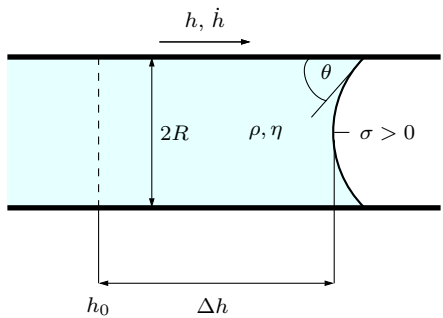
\begin{figure}[hb]
\centering{
\begin{tikzpicture}[>={Triangle[length=0pt 9,width=0pt 5]}]
\begin{scope}[rotate = -90]
\draw [color=MyBlue, fill=MyBlue] (2,-0.2)  -- (2,4.5) to [out=-45, in=-135] (4,4.5) -- (4,-0.2) --(2,-0.2);

\draw[ line width=2pt] (2,-0.2) -- (2, 5.5);
\draw[ line width=2pt] (4,-0.2) -- (4, 5.5);

    \draw[-, thick] (2,4.5) to[out=-45, in=-135] (4,4.5);
    
  \draw[<->] (2,1.7) to (4, 1.7);
  \draw[-] (3,4.1) to (3, 4.3);
    \draw[-] (3.1,4.1) to (5, 4.1);
      \draw[<->] (5,4.1) to (5, 1);

 \draw[->] (1.8,1.8) to (1.8, 3);
  \draw[->] (1.8,1.8) to (1.8, 3);
  \draw[dashed] (2,1) -- (4,1);
    \draw[-] (4,1) -- (5,1);
    \draw[-] (2,4.5) -- (2.8,3.8);
     \draw[-] (2,3.8) to[out=-60, in=-100] (2.58,4);

\node (C) at (3,3.2)  {\scriptsize $\rho, \eta$};  
\node (D) at (3,4.8)  {\scriptsize $\sigma>0$};  
\node (A) at (3,2)  {\scriptsize $2R$};
\node (E) at (5.3,2.5)  {\scriptsize $\Delta h$};
\node (F) at (5.3,1)  {\scriptsize $h_0$};
\node (G) at (2.2, 4)  {\scriptsize $\theta$};
\node (B) at (1.5,2.4)  {\scriptsize $h$, $\dot{h}$};
  
\end{scope}
\end{tikzpicture}}
    \caption{Sketch of the problem: meniscus traveling in a cylindrical tube.}\label{fig:notation}
\end{figure}

\paragraph{Mathematical formulation (see Fig. \ref{fig:notation}):} We consider a Newtonian liquid of density $\rho$ and viscosity $\eta$ inside a cylindrical tube of radius $R$. The surface tension between the liquid and the gas phase is denoted $\sigma$ and gravitational forces are assumed to be negligible. The equilibrium contact angle is denoted as $\theta_0$. Our primary quantity of interest is the "length" or the "height" of the meniscus (measured at the apex) inside the tube. We shall denote this height at time $t$ by $h(t)$. In physical units, the model equation reads as
\begin{align}\label{eqn:governing-equation}
2 \pi R \sigma \cos \theta_0 =8 \pi \eta h \dot{h}+\frac{d}{d t}\left(\pi R^2 \rho h \dot{h}\right) + 2 \pi R \zeta \dot{h}.
\end{align}
In fact, equation \eqref{eqn:governing-equation} expresses (under simplifying assumptions) a balance between the capillary driving force (provided that $\theta_0 < \pi/2$) with the viscous friction generated by the Hagen-Poiseuille flow away from the interface, the inertial term, and a contact line friction force that was introduced later by Martic et al.\ \cite{Martic2002}. The contact line friction parameter $\zeta > 0$ has units of viscosity and can be estimated from the Molecular Kinetic Theory of wetting (see, e.g., \cite{Blake2015,Blake1969}). The classical model is recovered for $\zeta=0$. Notice that recently, Delannoy et al. \cite{Delannoy2019} and Fricke et al.\  \cite{PreprintFricke2023a} showed that also the viscous dissipation generated by the flow close to the moving contact line (i.e.\ the "hydrodynamic theory") leads to an additional term with similar mathematical structure ($\sim 2 \pi R \ln(R/L) \dot{h}$). For simplicity, we stick to the above formulation, keeping in mind that it is possible to handle also other dissipative mechanisms.\\
\\
Note that the contact line friction force in \eqref{eqn:governing-equation} is appropriate for sufficiently small velocities of the contact line, i.e.\ for sufficiently small imbibition speed $\dot{h}$. For high velocities, equation \eqref{eqn:governing-equation} should be generalized to account for the non-linear dissipation at the contact line \cite{Martic2002}:
\begin{align*}
2 \pi R \sigma \left( \cos \theta_0 - c_1 \text{arcsinh}(c_2 \dot{h}) \right) =8 \pi \eta h \dot{h}+\frac{d}{d t}\left(\pi R^2 \rho h \dot{h}\right).
\end{align*}
However, we concentrate in the following on the small velocity regime described by \eqref{eqn:governing-equation}.\\
\\
It is convenient to introduce the following non-dimensional variables for the imbibition height and the time:
\[ H:=h/R, \quad s := t/(4 \eta R/\sigma). \]
Then, equation \eqref{eqn:governing-equation} takes the form
\begin{align}\label{eqn:governing-equation-nondim}
\cos \theta_0 = H(s) H'(s) + \frac{1}{32 \Oh^2} [H(s) H'(s)]' + \beta H'(s),
\end{align}
where the Ohnesorge number $\text{Oh}$ and the parameter $\beta > 0$ are defined as
\[ \Oh = \frac{\eta}{\sqrt{R \rho \sigma}} \quad \text{and} \quad \beta = \frac{\zeta}{4 \eta}. \]
Please notice that, for the remainder of this paper, we will assume that the viscosity is sufficiently high such that the inertial term $(HH')'/(32 \Oh^2)$ is negligible. So we concentrate on the governing equation of liquid imbibition in the simple form
\begin{align}\label{eqn:viscous-model-nondim}
\boxed{\cos \theta_0 = HH' + \beta H'.}
\end{align}
It is now an almost trivial exercise to compute the solution for the case $\beta=0$. The solution is the well-known $h \sim \sqrt{t}$ law, known as the "Washburn law" in the literature. More precisely, we have (for initial condition $H(0)=0$)
\[ H(s) = \sqrt{2 \cos(\theta_0) s} \quad \Leftrightarrow \quad  h(t)=\sqrt{\frac{R \sigma \cos \theta_0}{2 \eta} t}. \]
Notably, the problem can also be solved analytically for $\beta > 0$. And, moreover, this remains valid even if the pore is heterogeneous (see below).
\subsection{The Cassie-Baxter equation}\label{sec:1-2-cassie-baxter}
\paragraph{Static wetting on homogeneous surfaces:} The static contact angle $\theta_0$ of a liquid droplet in contact with a perfectly flat and homogeneous solid surface satisfies the famous Young equation \cite{Young1805}
\begin{align}\label{eqn:young}
\sigma \cos \theta_0 + \sigma_w = 0.
\end{align}
Here $\sigma > 0$ denotes the surface tension between the liquid and the gas phase and $\sigma_w$ is the specific surface energy for wetting the solid surface (notice that $\sigma_w < 0$ for $\theta_0 < \pi/2$). So the total surface energy reads as
\[ \mathcal{E} = \sigma A_{lg} + \sigma_w A_{ls}. \]
Here $A_{lg}$ and $A_{ls}$ are the total surface area of the liquid-gas and liquid-solid interface, respectively. Notice that, in the absence of gravity, equation \eqref{eqn:young} can easily be derived by minimizing the total surface energy under the constraint of volume conservation for the drop (see \cite{Gennes2004} and \cite{Fricke2021} for more details).

\paragraph{Static wetting on heterogeneous surfaces:}
Let us now recap the case of a perfectly planar but chemically heterogeneous surface composed of two different solid materials (labeled as "A" and "B") with distinct wettability. Cassie and Baxter \cite{Cassie1944,Cassie1948} studied this case with the aim of predicting the equilibrium contact angle of a liquid droplet on such a surface, again using energy minimization arguments. Let us assume that the characteristic length scale of the droplet is much larger than the characteristic length scale of the surface heterogeneity, i.e.\ the scale at which the surface chemistry changes. Let us further denote by $\theta_A$ and $\theta_B$ the equilibrium contact angle that the same liquid droplet forms on a planar surface of \text{only} \textit{A} or \textit{B}, respectively. Clearly, these contact angles satisfy the respective Young equations
\begin{align}\label{eqn:young-ab}
\sigma \cos \theta_A + \sigma_w^A = 0, \quad  \sigma \cos \theta_B + \sigma_w^B = 0.
\end{align}
Here, $\sigma_w^A$ and $\sigma_w^B$ denote the specific surface energies of wetting for the pairing $A$ and $B$ with the solid surface, respectively.\\
\\
We further assume that a portion $0 \leq r_A \leq 1$ of the surface is covered with material \text{A} and the remainder $r_B = 1 - r_A$ of the surface is covered with material \textit{B}. Now, the main assumption is that the motion of the contact line will lead to a change of surface energy that is simply a weighted average of the effects of the two materials with respect to $r_A$ and $r_B$. In other words, we assume that the specific wetting energy for the heterogeneous surface can effectively be approximated as
\begin{align}\label{eqn:effective_wetting_energy}
\sigma_w^{A,B} = r_A \sigma_w^A + r_B \sigma_w^B.
\end{align}
For this approximation to make sense, it is important to have the separation of scales described above. Then, any "significant" motion of the contact line will "see" a large enough statistical sample of the two materials. Finally, the Cassie-Baxter equation (for a perfectly planar surface) follows directly from \eqref{eqn:young} and \eqref{eqn:effective_wetting_energy}:
\begin{equation}\label{eqn:cassie-baxter}
\cos \theta_{\mathrm{CB}}=r_A \cos \theta_A+r_B \cos \theta_B.
\end{equation}
\paragraph{The CB-equation and dynamic wetting:} While the Cassie-Baxter (CB) equation was originally derived for \emph{static} wetting, it is often applied to \emph{dynamic} situations where a moving contact line interacts with a chemically or physically heterogeneous surface. In the context of imbibition into heterogeneous porous materials, the Cassie-Baxter contact angle is frequently combined with the Washburn equation to describe the influence of surface heterogeneity on the capillary-driven flow dynamics. For example, Kammerhofer et al.\ \cite{Kammerhofer2018} studied the interesting example of a powder bed composed of glass beads with different surface chemistry. The goal of this study is to better understand the reconstitution of food powders, where capillary-driven imbibition into heterogeneous porous structures plays a critical role. The authors studied the rise of water in the glass powder bed with different mixture fractions of the two materials and proposed the following adaptation of the Lucas-Washburn equation:
\begin{align}
h(t) = \sqrt{\frac{R_{\text{eff}} \sigma \cos \theta_{\mathrm{CB}} \, t}{2 \eta}}.
\end{align}
Here $R_{\text{eff}}$ is an effective pore radius that accounts effectively for the complex geometry of the pore spaces inside the powder bed and $\theta_{\mathrm{CB}}$ is the Cassie-Baxter angle defined by \eqref{eqn:cassie-baxter}. This approach caught our interest from a fundamental point of view and we decided to revisit the problem in a significantly simpler model system.

\clearpage
\section{The Lucas-Washburn equation for a heterogeneous material}\label{sec:2-analytical-solution}
In this section, we extend the classical Lucas-Washburn model to study capillary flow in a cylindrical pore with spatially varying wettability. Specifically, we assume that the equilibrium contact angle $\theta_0$ depends on the axial position of the contact line along the pore. Such a scenario models a heterogeneous surface where wettability changes either continuously or in a piecewise fashion, providing a simplified framework to explore the role of material heterogeneity on dynamic imbibition.\\
\\
We stick to the model problem \eqref{eqn:viscous-model-nondim}, where inertial effects are assumed to be negligible. Notice that we should also expect that the parameter $\beta = \zeta /(4 \eta)$ depends on the material. In fact, Blake et al.\ showed that the contact line friction $\zeta$ is strongly dependent on the work of adhesion \cite{Blake2004,Blake2015}. So mathematically, we study the following variant of the Lucas-Washburn equation
\begin{align}\label{eqn:viscous-model-nondim-heterogeneous}
\cos \theta_0(H) = HH' + \beta(H) H'.
\end{align}
For simplicity, we focus on a binary system where the wettability alternates between two distinct materials, A and B, characterized by $\theta_A$ and $\theta_B$ with $\cos \theta_A > \cos \theta_B > 0$. Clearly, this assumption can easily be dropped to investigate more complex cases. The key question we address is whether the \emph{Cassie-Baxter contact angle}, defined as the weighted average of the local cosines of the contact angles,
\[
\cos \theta_{\mathrm{CB}} = r_A \cos \theta_A + r_B \cos \theta_B,
\]
remains a valid descriptor for the dynamic case. To this end, we compute the meniscus travel time, or \emph{crossing time}, through a heterogeneous pore, and compare it to the prediction based on an effective Cassie-Baxter angle. Our model problem provides a clear framework for identifying deviations from the Cassie-Baxter theory in dynamic situations and elucidates the role of the spatial arrangement of materials in optimizing fluid transport. We shall now derive the analytical solution for this system and analyze its implications for random and ordered material distributions.

\subsection{Analytical solution for the heterogeneous case}
The problem can directly be solved using the method of separation of variables. We observe, that we can rewrite \eqref{eqn:viscous-model-nondim-heterogeneous} in the form
\begin{align}\label{eqn:viscous-model-nondim-heterogeneous-v2}
\frac{dH}{ds} = \frac{\cos \theta_0(H)}{H + \beta(H)}.
\end{align}
Hence, the dimensionless time $s$ to travel from height $0$ to a certain height $H$ is given as
\begin{align}\label{eqn:solution-integral-form}
s = \int_0^H \frac{d \tilde{H}}{H'(\tilde{H})} = \int_0^H \frac{\tilde{H} + \beta(\tilde{H})}{\cos \theta_0(\tilde{H})} \, d \tilde{H}.
\end{align}
Mathematically, equation \eqref{eqn:solution-integral-form} is an integral representation of the inverse function $s=s(H)$ of the (monotonic!) solution $H=H(s)$ of \eqref{eqn:viscous-model-nondim-heterogeneous-v2}. This is a well-suited form of the solution since we are interested primarily in the crossing time.

\paragraph{Binary system:} Let us now focus on the case of a binary system with materials $A$ and $B$. Then, the pore is partitioned into a finite set of intervals (or "segments") $[H_i, H_{i+1}]$ such that $\cos \theta_0$ and $\beta$ are constant on each of them. We denote by $L_i = H_{i+1} - H_i$ and $\bar{H}_i = (H_{i+1}+H_i)/2$ the length and center position of segment $i$, respectively.\\
\begin{figure}[htbp]
\centering
\includegraphics[width=0.65\columnwidth]{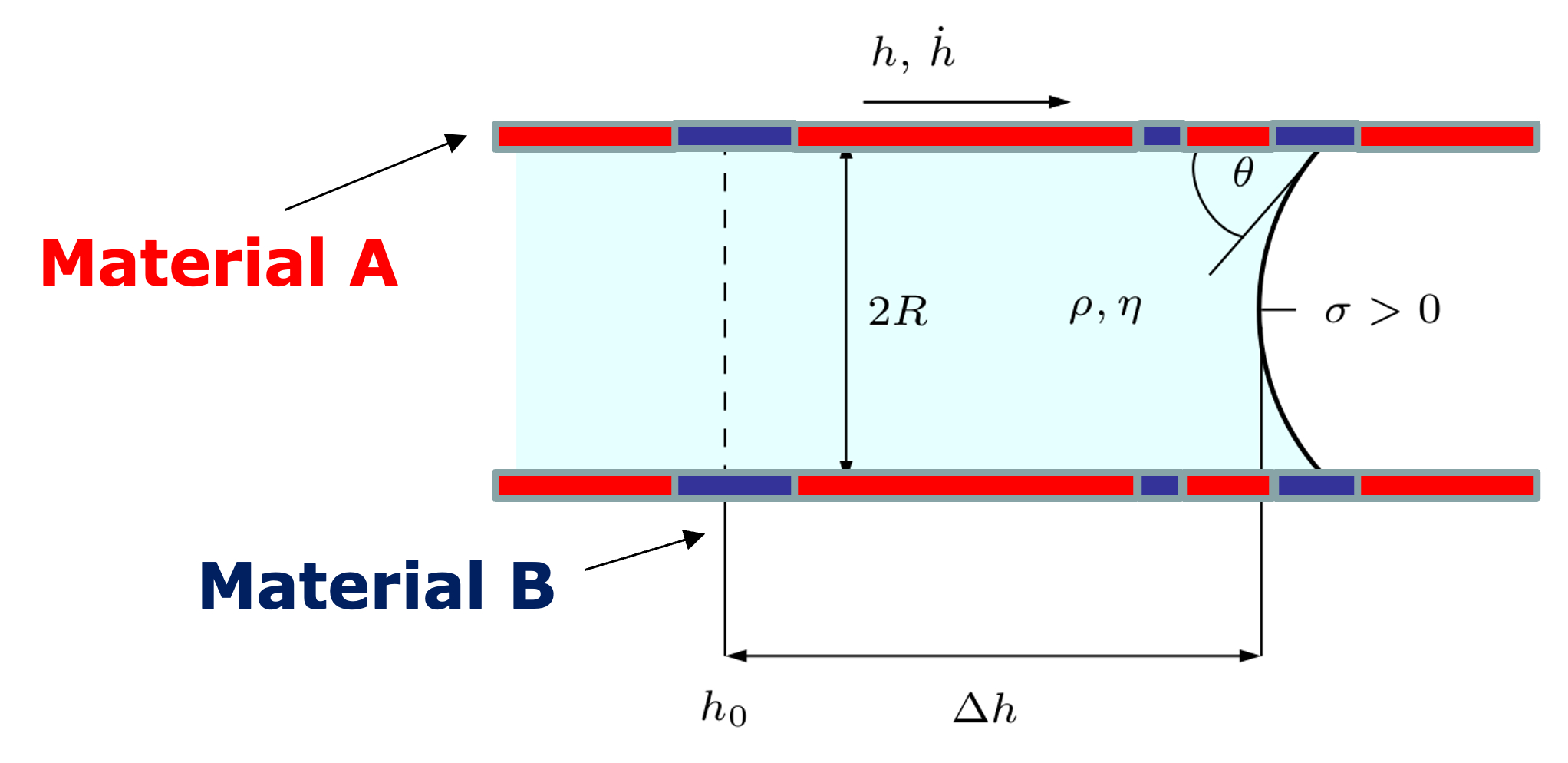}
\caption{Binary model system.}\label{fig:binary-model-system}
\end{figure}

Let us write $\theta_i$ and $\beta_i$ as a short notation for the contact angle and the (friction) parameter $\beta$ on the $i$-th interval. Then we can evaluate the integral in \eqref{eqn:solution-integral-form} to find\footnote{Notice that $(H_{i+1}^2-H_i^2)/2=(H_{i+1}-H_i)(H_{i+1}+H_i)/2 = L_i \bar{H}_i$.}:
\begin{equation}\label{eqn:crossing-time-solution-binary}
\begin{aligned}
s &= \sum_i \int_{H_i}^{H_{i+1}} \frac{\tilde{H} + \beta_i}{\cos \theta_i} \, d \tilde{H} = \sum_i \left( \frac{\beta_i}{\cos \theta_i} L_i + \frac{H_{i+1}^2-H_i^2}{2 \cos \theta_i} \right)\\
 &= \frac{\beta_A L_A}{\cos \theta_A} + \frac{\beta_B L_B}{\cos \theta_B} + \sum_i \frac{L_i \bar{H}_i}{\cos \theta_i} =: s_\beta + \sum_i \frac{L_i \bar{H}_i}{\cos \theta_i}.
\end{aligned}
\end{equation}

\subsection{The role of material distribution}
We can draw a couple of conclusions directly from the solution \eqref{eqn:crossing-time-solution-binary}.
\begin{enumerate}[(i)]
\item The crossing time $s$ has two components, namely the component $s_\beta$ due to the "friction parameter" $\beta$
\[ s_\beta = \frac{\beta_A L_A}{\cos \theta_A} + \frac{\beta_B L_B}{\cos \theta_B}, \]
and the component $s_P$ due to the viscous friction within the Hagen-Poiseuille flow
\begin{align}\label{eqn:sp-definition}
s_P := \sum_i \frac{L_i \bar{H}_i}{\cos \theta_i}.
\end{align}
\item Obviously, for a homogeneous material, we have $\bar{H} = H/2$ and we recover the well-known Lucas-Washburn law (neglecting $s_\beta$):
$$s_P = H^2/(2 \cos \theta_0) \quad \text{or} \quad H = \sqrt{2 \cos(\theta_0) \, s_P}.$$
\item Crucially, the component $s_\beta$ is \emph{invariant} with respect to a redistribution of the material within the tube, but the component $s_P$ is \emph{not}. So, in fact, the component $s_P$ is sensitive to the order of the material within the tube.
\end{enumerate}

\subsection{Ordered states - hydrophobic-hydrophilic vs.\ hydrophilic-hydrophobic}
\begin{figure}[htbp]
\centering
\subfigure[More hydrophobic material $B$ first (BA).]{\includegraphics[width=0.4\columnwidth]{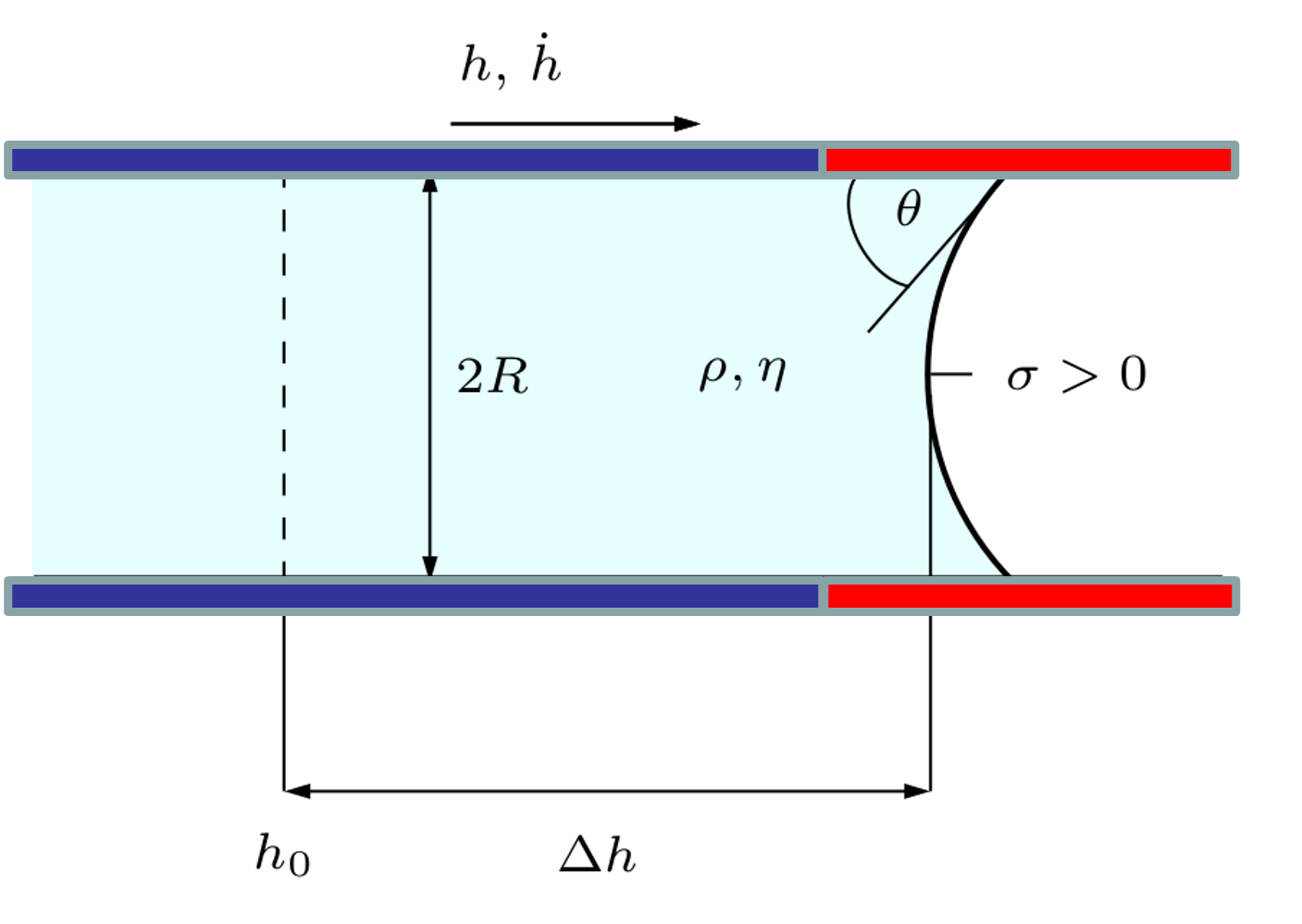}\label{fig:hydrophobic-first}}
\subfigure[Hydrophilic material $A$ first (AB).]{\includegraphics[width=0.4\columnwidth]{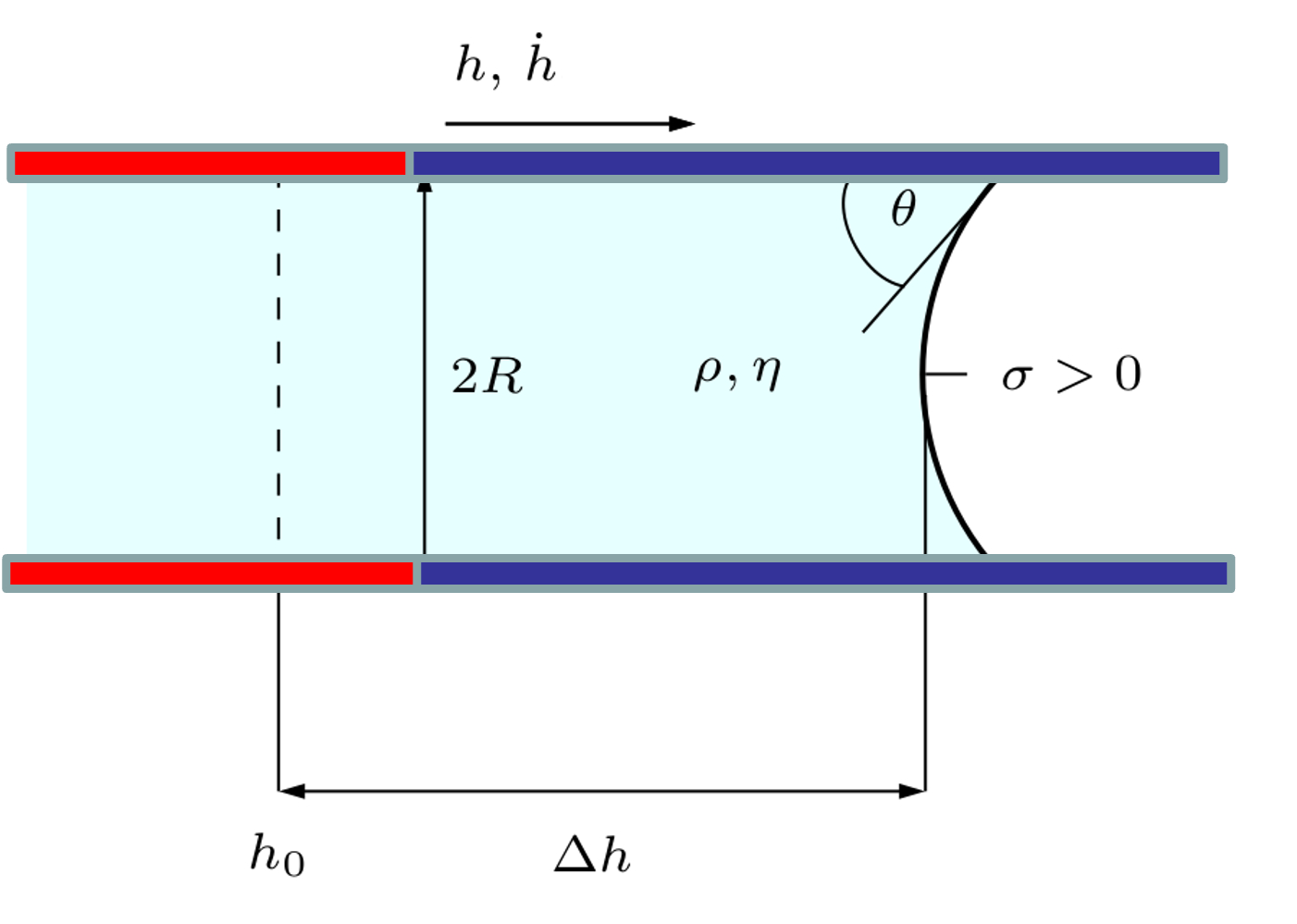}\label{fig:hydrophilic-first}}
\caption{Comparison of completely ordered states.}\label{fig:eat-the-frog-setup}
\end{figure}

In order to understand the basic phenomenon, we consider the limiting cases of a completely ordered binary system of materials $A$ (surface ratio $r_A$) and material $B$ (surface ratio $r_B=1-r_A$). We assume $\cos \theta_A > \cos \theta_B > 0$. Hence, both materials are hydrophilic but material $A$ (marked in red in Figures \ref{fig:binary-model-system} and \ref{fig:eat-the-frog-setup}) is more hydrophilic compared to $B$. In this sense, we call material $B$ (marked in blue in Figures \ref{fig:binary-model-system} and \ref{fig:eat-the-frog-setup}) "hydrophobic".\\
\\
We assume that the tube has the total (dimensionless) length $L > 0$. We consider the two limiting cases where all of $B$ is placed before $A$ (state "BA", see Fig. \ref{fig:hydrophobic-first}) or vice versa (state "AB", see Fig. \ref{fig:hydrophilic-first}). Hence, "BA" can be transformed to "AB" by simply flipping the tube. Thanks to the solution \eqref{eqn:crossing-time-solution-binary}, we can easily compare the crossing time for the two configurations. We know that $s_\beta$ is the same for both cases. Hence, we may concentrate on the component $s_P$. A simple calculation shows that:
\begin{mdframed}[linewidth=0.8pt, linecolor=black]
\begin{equation}\label{eqn:eat-the-frog}
\begin{aligned}
s_P^{AB} &= L^2 \left( \frac{r_A^2}{2 \cos \theta_A} + \frac{r_B^2 + 2 r_A r_B}{2 \cos \theta_B} \right) \\
s_P^{BA} &= L^2 \left( \frac{r_A^2 + 2 r_A r_B}{2 \cos \theta_A} + \frac{r_B^2}{2 \cos \theta_B} \right) \\
\Rightarrow \quad s_P^{AB} - s_P^{BA} &= L^2 r_A r_B \left( \frac{1}{\cos \theta_B} - \frac{1}{\cos \theta_A} \right)
\end{aligned}
\end{equation}
\end{mdframed}
As noted before, the crossing times for the two configurations are not equal because of the non-linearity of the problem. In particular, we observe from \eqref{eqn:eat-the-frog} that the "more hydrophobic first" (BA) configuration is physically preferred in the sense that the imbibition is faster:
\[ s_P^{AB} > s_P^{BA} \quad \text{if} \quad \cos \theta_A > \cos \theta_B > 0.  \]
In this sense, the meniscus in the heterogeneous Lucas-Washburn model prefers to "eat the frog", a concept introduced by productivity consultant Brian Tracy in his book \emph{Eat That Frog} \cite{Tracy2007}. Much like tackling the most challenging task first to optimize productivity, the system achieves faster imbibition by addressing the "harder-to-wet" (more hydrophobic) material at the beginning of the process.

\clearpage
\section{Results for random binary materials with a Gaussian distribution}\label{sec:3-binary-materials}
\begin{figure}[ht]
\centering
\subfigure[$l_A$]{\includegraphics[width=0.45\columnwidth]{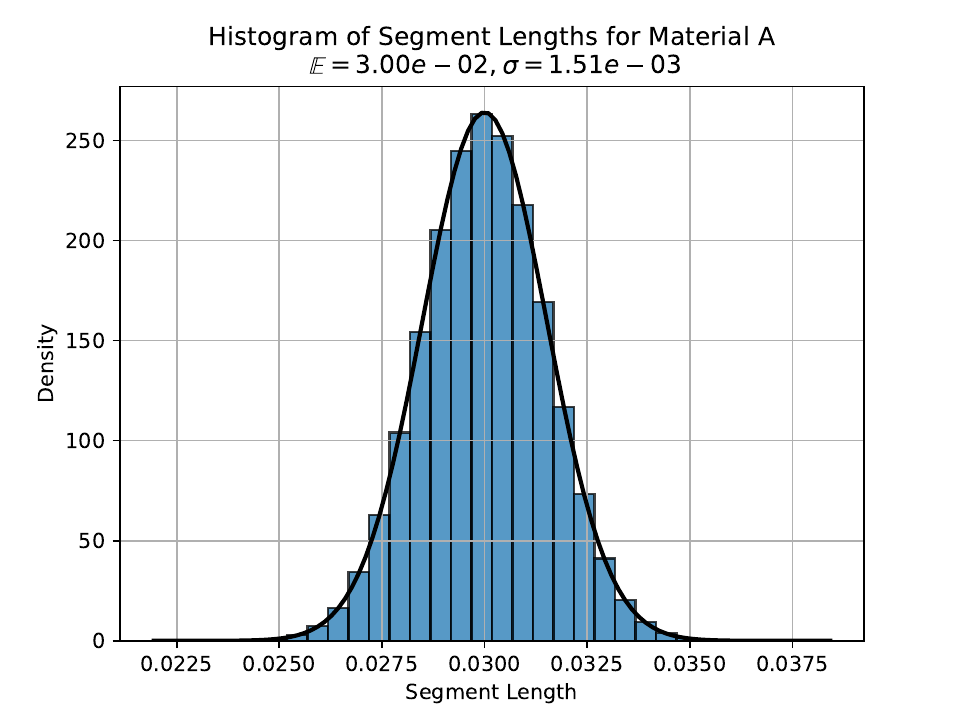}}
\subfigure[$l_B$]{\includegraphics[width=0.45\columnwidth]{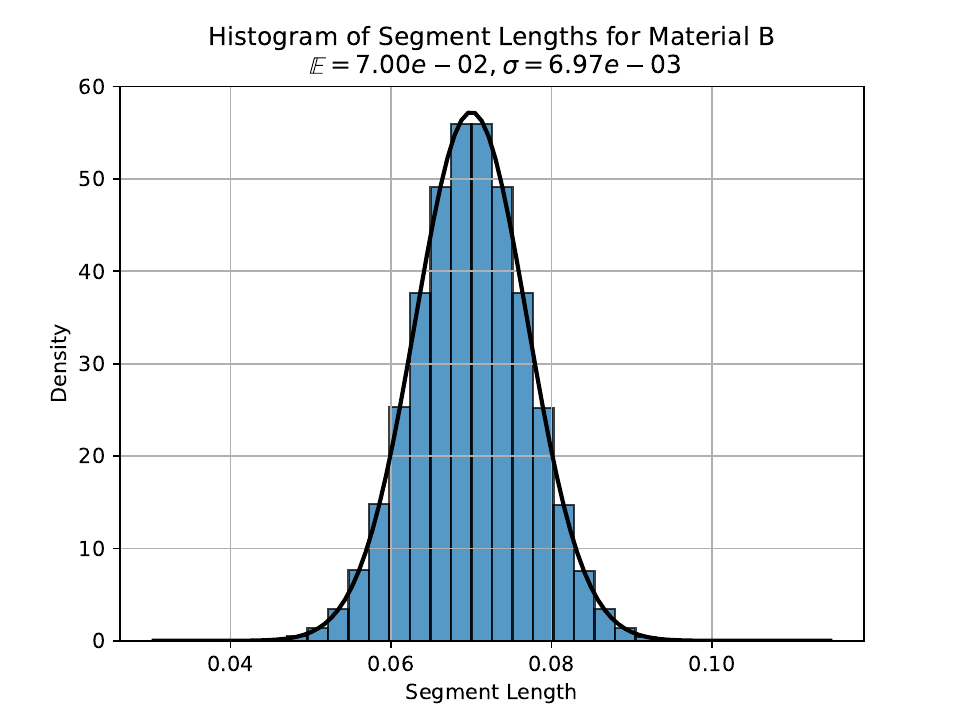}}
\caption{Empirical distributions of the segment lengths $l_A$ and $l_B$.}\label{fig:la_lb}
\end{figure}

In the previous section, we explored capillary imbibition in heterogeneous tubes with ordered distributions of materials A and B, revealing that the spatial arrangement significantly influences the crossing time. We now extend this analysis to the case of disordered heterogeneity, where the lengths of the material segments follow a normal (Gaussian) distribution. Such disordered systems provide a more realistic model for natural and engineered porous materials, where material heterogeneity arises due to randomness in surface properties or fabrication processes.\\
\\
The primary goal of this section is to investigate how statistical variations in segment lengths influence the imbibition dynamics and whether the Cassie-Baxter contact angle remains a valid descriptor in such a dynamic setting. By analyzing the crossing time for random distributions, we aim to uncover deviations from the predictions based on the effective contact angle and propose a more accurate averaging approach. We further explore the statistical distribution of crossing times as the number of segments increases, addressing the asymptotic behavior in the limit of large segment numbers.\\
\\
We implemented a simple code in Python \cite{Fricke2024-code} that creates samples of the binary system with total dimensionless length $L$ with certain statistical properties. The idea is to sample $\Nseg$ (where $\Nseg \in \N$ is a free parameter) segments of $A$ and $B$ from the (independently distributed) Gaussian distributions in an alternating way. After the sampling is completed, the total sample is rescaled uniformly to the length $L$.\\
\\
We denote by $l_A$ and $l_B$ the length of an individual segment of $A$ and $B$, respectively. The proportion of the two expected values $\mathbb{E}[l_A]$ and $\mathbb{E}[l_B]$ determines the expected value of the ratio of material $A$ according to
\begin{align}\label{eqn:r_A_proportion_formula}
\mathbb{E}[r_A] = \frac{\mathbb{E}[l_A]}{\mathbb{E}[l_A]+\mathbb{E}[l_B]} = \left( 1 + \frac{\mathbb{E}[l_B]}{\mathbb{E}[l_A]} \right)^{-1} \quad \Leftrightarrow \quad \frac{\mathbb{E}[l_B]}{\mathbb{E}[l_A]} = \frac{1-\mathbb{E}[r_A]}{\mathbb{E}[r_A]}.
\end{align}
In practice, we want to fix $\mathbb{E}[r_A]$ as a physical parameter. So we will use the target value of $\mathbb{E}[r_A]$ to determine the ratio $\mathbb{E}[l_B]/\mathbb{E}[l_A]$. Since we apply a normalization step in the end, we may also assume for simplicity that $\mathbb{E}[l_A] = 1$. This simplifies the implementation. Moreover, the user can specify the relative standard deviation (normalized by the expected value) of the segment length, i.e.\
\begin{align*}
\hat\sigma[l_A] := \sigma[l_A]/\mathbb{E}[l_A], \quad \text{and} \quad \hat\sigma[l_B] := \sigma[l_B]/\mathbb{E}[l_B].
\end{align*}
After a material sample is created, we evaluate the viscous crossing time $s_P$ by a simple evaluation of the sum in \eqref{eqn:sp-definition}. This process is repeated $\Nsamp$-times and then the statistical distribution of $l_A$, $l_B$ and the viscous crossing time $s_P$ is investigated.
\begin{figure}[hb]
\centering
\subfigure[$l_A$ vs.\ $l_B$]{\includegraphics[width=0.45\columnwidth]{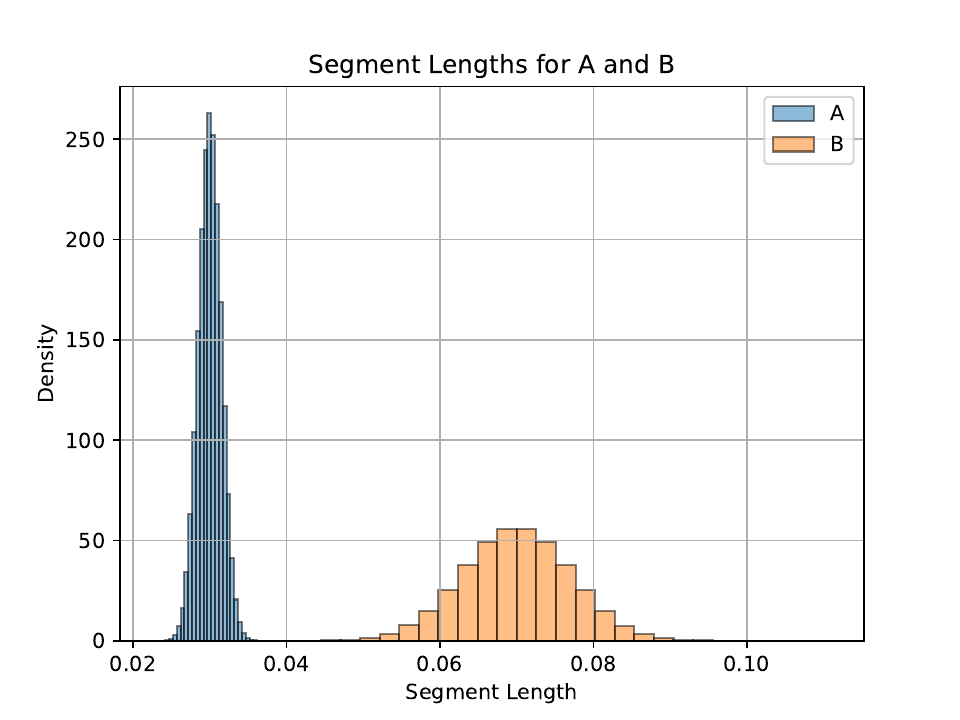}}
\subfigure[$r_A$]{\includegraphics[width=0.45\columnwidth]{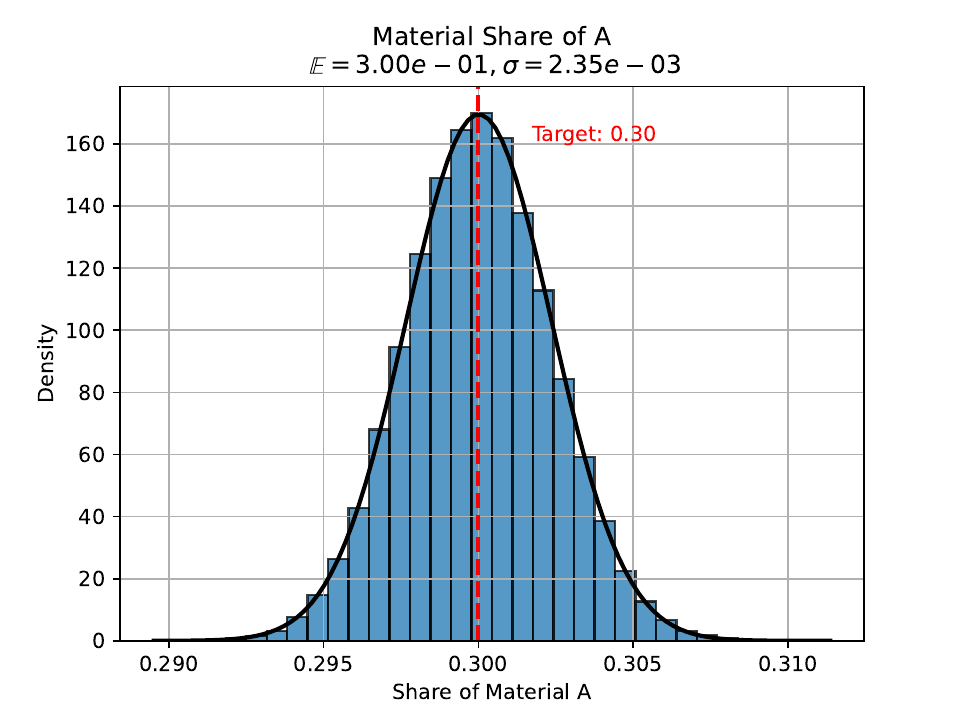}}
\caption{Comparison of $l_A$ and $l_B$, distribution of the share of A.}\label{fig:r_A}
\end{figure}

\subsection{A numerical example}
To illustrate the results, we choose one particular example with the following dimensionless physical parameters:
\begin{align}\label{eqn:parameters-1}
\theta_A = 30^\circ, \ \theta_B = 60^\circ, \quad \mathbb{E}[r_A] = 0.3, \quad \hat\sigma[l_A]  = 0.05, \ \hat\sigma[l_B]  = 0.1, \quad L = 10, \quad \Nseg = 100.
\end{align}
Please note, that the findings discussed below are general and do not depend on the above choice of parameters. We choose $\Nsamp = 10^5$ and verify that the resulting distributions of $l_A$, $l_B$ and the viscous crossing time $s_P$ are Gaussian as well. The full implementation in Python to generate the following results can be found here \cite{Fricke2024-code}. The implementation allows to easily repeat the study with different parameters.

\paragraph{Statistics of the segment length and material share:} We run the Python code over $\Nsamp = 10^5$ samples with the parameters \eqref{eqn:parameters-1} and report the statistics of $l_A$, $l_B$ and $r_A$; see Figures \ref{fig:la_lb} and \ref{fig:r_A}. A Gaussian distribution is fitted to the empirically collected histograms confirming that the distributions of $l_A$, $l_B$ are indeed Gaussian with the predefined relative standard deviations. As a result, the share of material $A$ is also Gaussian distributed with
\[ \mathbb{E}[r_A] = 0.3 \quad \text{and} \quad \sigma[r_A] = 2.35 \cdot 10^{-3}.  \]

\begin{figure}
\centering
\includegraphics[width=0.6\columnwidth]{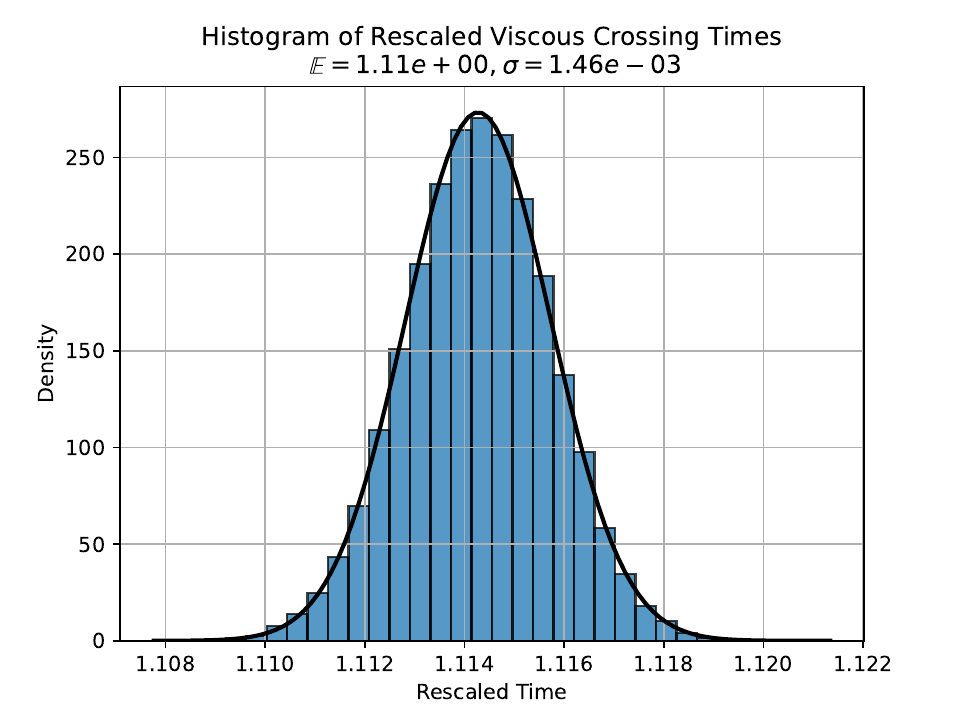}
\caption{Empirical distribution of the viscous crossing time $s_P$ normalized with respect to the optimum $s_P^{BA}$.}\label{fig:crossing_time_distribution}
\end{figure}

\paragraph{Statistics of the crossing time:} Figure \ref{fig:crossing_time_distribution} shows the distribution of $s_P$, i.e.\ the viscous crossing time rescaled by the optimum $s_P^{BA}$. The distribution is well-approximated by a Gaussian distribution with
\[ \mathbb{E}[s_P/s_P^{BA}] \approx 1.11, \quad \text{and} \quad  \sigma[s_P/s_P^{BA}] \approx 1.45 \cdot 10^{-3}. \]
Hence, the expected value of the viscous crossing time is approximately $11\%$ higher than the optimum (realized for the ordered $BA$-state).

\subsection{An effective contact angle model to predict the viscous crossing-time}
\begin{figure}[hb]
\centering
\includegraphics[width=0.6\columnwidth]{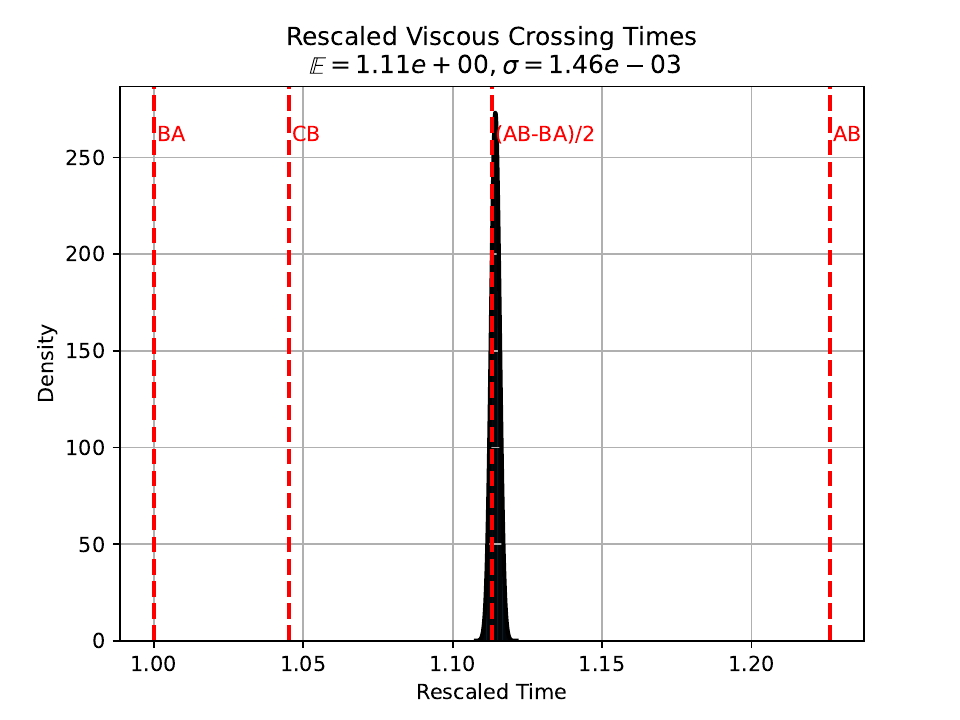}
\caption{Distribution of $s_P/s_P^{BA}$ compared with reference values.}\label{fig:crossing_time_distribution_references}
\end{figure}

To summarize, we analyzed the statistical behavior of the viscous crossing time for random binary systems where segment lengths follow a Gaussian distribution. While the Lucas-Washburn equation provides a direct means of computing the dynamics in such systems, it remains unclear whether the Cassie-Baxter contact angle, originally derived for static wetting, can effectively predict the dynamic imbibition behavior in heterogeneous materials. Therefore, we address the following central question:\\
\\
\emph{Is the Cassie-Baxter contact angle a reliable descriptor for the dynamics of the capillary-driven imbibition in our disordered heterogeneous model system?}\\
\\
To this end, we compare the statistical results for the viscous crossing time against predictions based on the Cassie-Baxter angle. We can readily compute the Cassie-Baxter angle (defined with respect to $\mathbb{E}[r_A]$) from the parameters defined in \eqref{eqn:parameters-1} leading to
\begin{align}
\cos \theta_{\mathrm{CB}}= 0.3  \cos(30^\circ)+ 0.7 \cos(60^\circ) \approx 0.610 \quad \Rightarrow \quad \theta_{\mathrm{CB}} \approx 52.42^\circ.
\end{align}
We define the "Cassie-Baxter viscous crossing time" as the viscous crossing time for a homogeneous tube of the same dimensionless length $L$ and uniform contact angle $\theta_{\mathrm{CB}}$:
\begin{align}
s_P^{\mathrm{CB}} := \frac{L^2}{2 \cos \theta_{\mathrm{CB}}} = \frac{L^2}{2 (r_A \cos \theta_A + r_B \cos \theta_B)}.
\end{align}
Figure \ref{fig:crossing_time_distribution_references} shows a comparison of the empirical distribution viscous crossing time compared with $s_P^{BA}$ (best case), $s_P^{AB}$ (worst case) and the Cassie-Baxter prediction $s_P^{\mathrm{CB}}$. Notably, the observed crossing time is significantly off from the Cassie-Baxter prediction:
\[  s_P/s_P^{BA} = 1.11 \pm 0.00145, \quad s_P^{\mathrm{CB}}/s_P^{BA} \approx 1.05.  \]
This is a significant result because it shows that one must be careful in passing from a static (Cassie-Baxter equation) to a dynamic problem (capillary imbibition in the viscous regime).

\begin{figure}[htb]
\centering
\includegraphics[width=0.6\columnwidth]{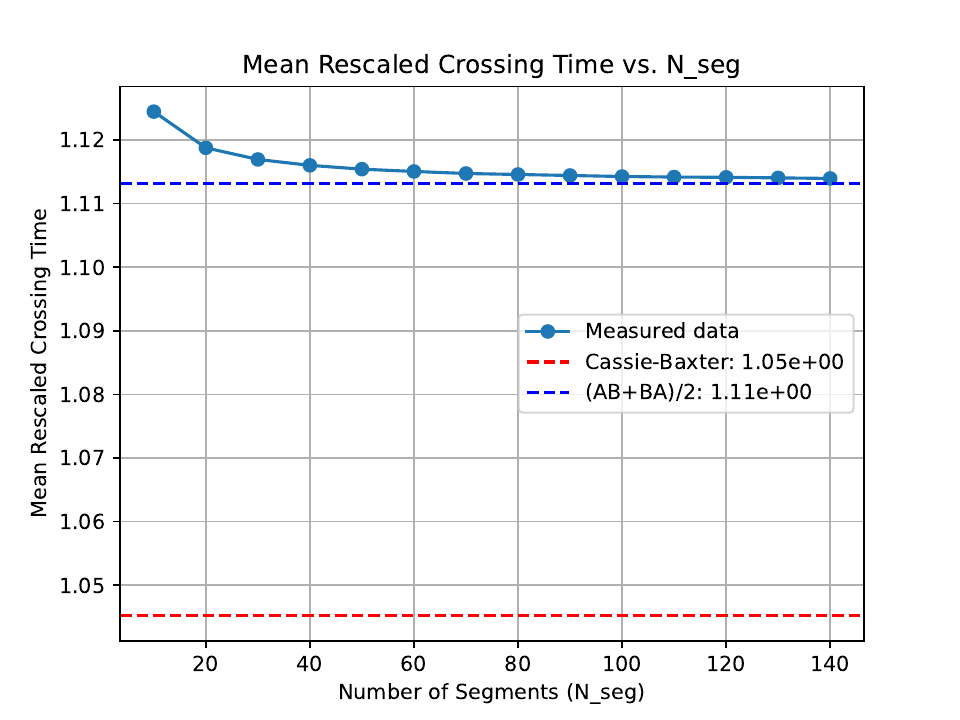}
\caption{Behavior of $\mathbb{E}[s_P/s_P^{BA};\Nseg]$ as $\Nseg \rightarrow \infty$.}\label{fig:nseg_study}
\end{figure}

\paragraph{Asymptotic limit for $\Nseg \rightarrow \infty$:} One may wonder if the number of segments $\Nseg$ plays a crucial role for the phenomenon described above. To answer this question, we performed a parameter study in $\Nseg$. The number of material boundaries that the contact line has to pass goes to infinity as the number of segments (of both $A$ and $B$) grows for a fixed ratio $\mathbb{E}[r_A]$. This can be seen as a "full homogenization" of the material in the limit.
It is not immediately obvious whether an asymptotic limit exists and what form it takes. Figure \ref{fig:nseg_study} shows the expected value $\mathbb{E}[s_P/s_P^{BA}]$ as a function of $\Nseg$ (for $\Nsamp = 10^5$). We observe that, in fact, the expected value converges to the arithmetic mean of the best and the worst case, i.e.\
\begin{align}
\lim_{\Nseg \rightarrow \infty} \mathbb{E}[s_P;\Nseg] = \frac{s_P^{BA}+s_P^{AB}}{2}.
\end{align}
Interestingly, this limiting case can be understood analytically. To this end, we compute the viscous crossing time for the case of $\Nseg$ segments of $A$ and $B$ with fixed ratio $r_A$ and \emph{vanishing} standard deviations $\sigma[l_A]$ and $\sigma[l_B]$. Hence, we assume that all the $A$ and $B$ segments are of the same length, respectively. Then we can write down the crossing time for such a configuration as a function of $\Nseg$.\\
\\
Since $l_A$ and $l_B$ are constant, we have
\[ l_A + l_B = L/\Nseg. \]
Moreover, the proportion of $l_A$ and $l_B$ can be determined from $r_A$ using the relation \eqref{eqn:r_A_proportion_formula}. From these two conditions, we obtain complete information on $l_A$ and $l_B$:
\begin{align*}
l_A = r_A \, \frac{L}{\Nseg}, \quad l_B = (1-r_A) \, \frac{L}{\Nseg}.
\end{align*}
Now we are in the position to compute $s_P$. Let us denote by $I_A$ and $I_B$ the set of all the indices of segments of material $A$ and $B$, respectively. Then, the viscous crossing time $s^\ast_P$ for this configuration (as a function of $\Nseg$) reads as:
\begin{equation}\label{eqn:asymptotic_limit_eq1}
\begin{aligned}
s^\ast_P(\Nseg) &= \sum_{i \in I_A} \frac{l_A \bar{H}_i}{\cos \theta_A} + \sum_{i \in I_B} \frac{l_B \bar{H}_i}{\cos \theta_B} \\
&= \frac{L}{\Nseg} \left( \frac{r_A}{\cos \theta_A} \sum_{i \in I_A} \bar{H}_i + \frac{1-r_A}{\cos \theta_B} \sum_{i \in I_B} \bar{H}_i \right) \\
&= L \left( \frac{r_A}{\cos \theta_A} \left[ \frac{1}{\Nseg} \sum_{i \in I_A} \bar{H}_i \right] + \frac{1-r_A}{\cos \theta_B} \left[ \frac{1}{\Nseg} \sum_{i \in I_B} \bar{H}_i \right]   \right)
\end{aligned}
\end{equation}
Now we notice that the terms
\[ \frac{1}{\Nseg} \sum_{i \in I_A} \bar{H}_i, \quad \frac{1}{\Nseg} \sum_{i \in I_B} \bar{H}_i \]
are nothing but an average over all the positions of segments with material $A$ and $B$, respectively. Both terms converge to the limit $L/2$ as $\Nseg \rightarrow \infty$ since the segments are distributed uniformly over the interval $[0,L]$. Hence it follows directly from \eqref{eqn:asymptotic_limit_eq1} that
\begin{align}\label{eqn:asymptotic_limit_eq2}
s_P^\ast := \lim_{\Nseg \rightarrow \infty} s^\ast_P(\Nseg) = \frac{L^2}{2} \left( \frac{r_A}{\cos \theta_A} + \frac{r_B}{\cos \theta_B} \right).
\end{align}
A short calculation shows that, indeed, the asymptotic limit $s_P^\ast$ is nothing but the arithmetic average between the best and the worst case:
\[ s_P^\ast  = \frac{L^2}{2} \left( \frac{r_A}{\cos \theta_A} + \frac{r_B}{\cos \theta_B} \right) = \frac{s_P^{AB}+s_P^{BA}}{2}. \]

\begin{figure}[hbt]
\centering
\includegraphics[width=0.65\columnwidth]{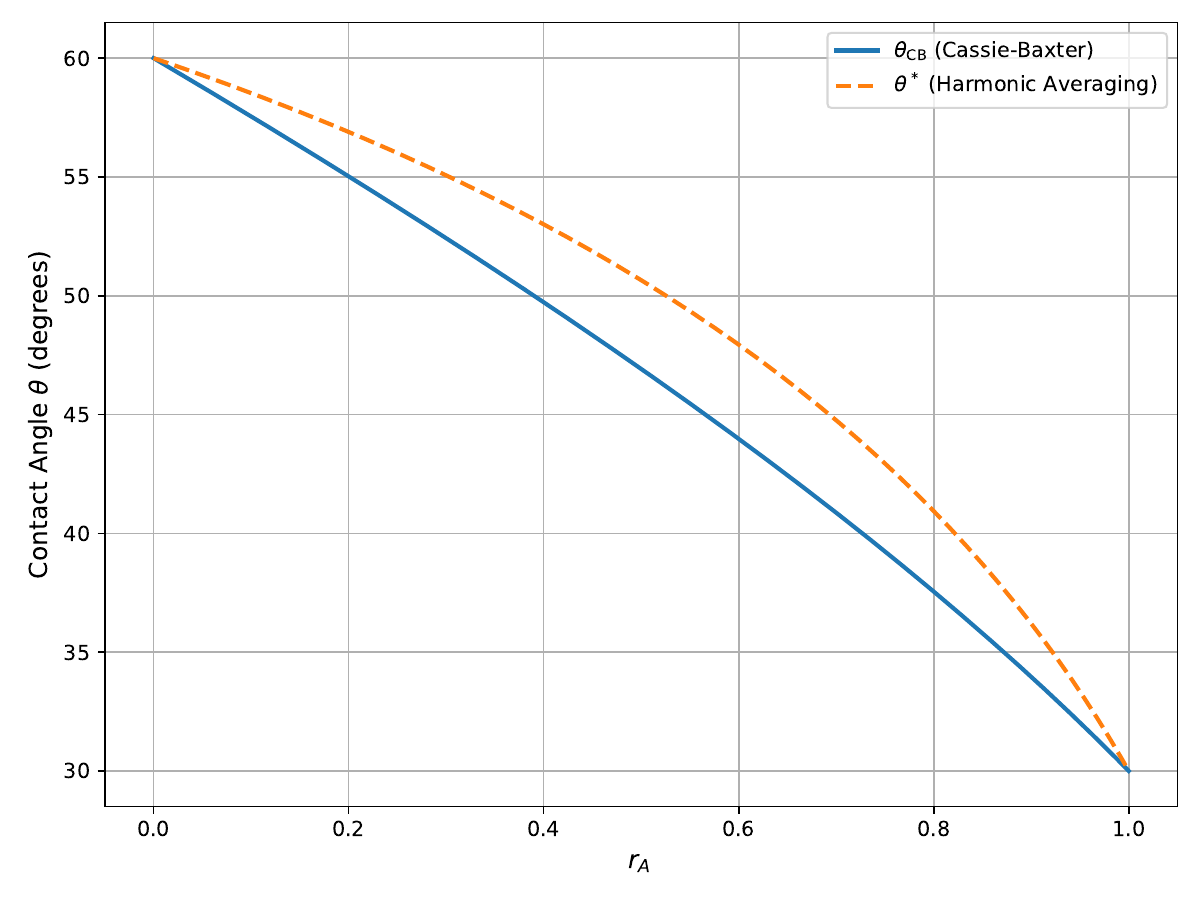}
\caption{Comparison of $\theta_{\mathrm{CB}}$ and $\theta^\ast$ as a function of the material ratio $r_A$.}\label{fig:averaging}
\end{figure}

\paragraph{Definition of the effective contact angle $\theta^\ast$:} So now we can simply read off the appropriate definition of the averaged contact angle $\cos \theta^\ast$ by identifying $s_P^\ast$ in \eqref{eqn:asymptotic_limit_eq2} with the crossing time of a homogeneous tube:
\[ s_P^\ast = \frac{L^2}{2} \left( \frac{r_A}{\cos \theta_A} + \frac{r_B}{\cos \theta_B} \right) \stackrel{!}{=} \frac{L^2}{2 \cos \theta^\ast}.  \]
The result is
\begin{equation}\label{eqn:new-contact-angle-averaging}
\boxed{\cos \theta^\ast = \left(  \frac{r_A}{\cos \theta_A}  +  \frac{r_B}{\cos \theta_B} \right)^{-1}.}
\end{equation}
This result contrasts with the classical Cassie-Baxter contact angle
\[ \cos \theta_{\mathrm{CB}}=r_A \cos \theta_A+r_B \cos \theta_B. \]
This shows that, instead of the classical Cassie-Baxter contact angle, the weighted harmonic averaging defined in \eqref{eqn:new-contact-angle-averaging} describes the viscous crossing time under the given assumptions. For the parameters given in \eqref{eqn:parameters-1}, we find
\[ \theta^\ast \approx 55.07^\circ \quad > \quad \theta_{\mathrm{CB}} \approx 52.42^\circ. \]
Even though the difference between the two models ($2.5^\circ$ degrees in this examples) seems minor, this non-linear averaging principle is very interesting from a fundamental research point of view; see Fig. \ref{fig:averaging}. It should be investigated in more detail in the future.
\clearpage
\section{Conclusion}\label{sec:4-conclusion}
In this work, we demonstrated a non-trivial dependence of the imbibition speed on the spatial distribution of wettability within a heterogeneous material. Starting from the Lucas-Washburn equation, we derived analytical solutions for ordered and disordered distributions of two materials and showed that the arrangement of wettability has a significant impact on the viscous crossing time. For ordered systems, we identified the "more hydrophobic-first" configuration as optimal, while for disordered systems, we proposed a new \emph{weighted harmonic averaging} method to predict the effective contact angle, which accurately captures the observed dynamics. To the best of our knowledge, this is the first time that the relationship between spatial wettability heterogeneity and imbibition dynamics has been explicitly demonstrated in this form.\\
\\
Several avenues for further research are identified:
\begin{itemize}
    \item \textbf{Relaxing axial symmetry assumptions}: Future studies should investigate systems where axial symmetry is broken, as this may introduce additional complexities in the capillary flow dynamics.
    \item \textbf{Contact angle relaxation}: The relaxation of the contact angle when the contact line crosses a wettability boundary could play a critical role in the dynamics and should be explored using Computational Fluid Dynamics (CFD) simulations.
    \item \textbf{Experimental validation}: Experimental studies are essential to validate the theoretical predictions and quantify the effect of material heterogeneity in real systems.
\end{itemize}
Our findings provide new insights into the role of material heterogeneity in capillary-driven flows, paving the way for optimized design of porous materials in applications ranging from engineered surfaces to natural systems.

\subsection*{Acknowledgments}
We acknowledge the financial support by the German Research Foundation (DFG) within the Collaborative Research Centre 1194 (Project-ID 265191195). We further thank the organizers of the \emph{12th IFPRI Workshop on Particle Technology: Powder Reconstitution}, hosted by the International Fine Particle Research Institute in Toronto, Canada, in June 2024, for their kind invitation. The preparation of the keynote presentation by Mathis Fricke at this workshop provided the initial motivation for this investigation.

\bibliographystyle{plainurl}

\end{document}